\documentstyle[11pt,aaspp4]{article}

\slugcomment{Submitted to the Astrophysical Journal}

\lefthead{}
\righthead{}

\begin{document}

\title{Ultra High Energy Cosmic Rays and the Large Scale\\
       Structure of the Galactic Magnetic Field}
\author{Todor Stanev}
\affil{Bartol Research Institute, University of Delaware, Newark
DE 19711, U.S.A.}

\begin{abstract} 

  We study the deflection of ultra high energy cosmic ray protons
 in different models of the regular galactic magnetic field. Such
 particles have gyroradii well in excess of 1 kpc and their propagation
 in the Galaxy is not affected by the small scale structure of the
 field. Their trajectories, however, reflect the large scale
 structure. A future large experimental statistics of cosmic rays
 of energy above 10$^{19}$ eV could be used for a study of the
 large scale structure of the galactic magnetic field if such
 cosmic rays are indeed charged nuclei accelerated at powerful
 astrophysical objects and if the distribution
 of their sources is not fully isotropic.  

  We also show the corrections of the arrival directions
 due to the deflection in the galactic magnetic field of
 for  a subsample of the world statistics of cosmic rays of energy
 above 2$\times$10$^{19}$ eV and discuss the implications for
 the search for the sources of these events.
\end{abstract}

\section{Introduction}
 
  While the majority of cosmic rays detected at Earth is quite
 isotropic, an observable anisotropy appear to set up (\cite{Watson})
 at the approach of energies of order 10$^{18}$ eV (1~EeV). This effect
 is quite natural, because the gyroradius of protons of that
 energy in 1 $\mu$G magnetic field is about 1 kpc, i.e. of O(10)
 larger than the scale of the random component of the galactic
 magnetic field (GMF). Although the typical values of the regular
 component of the GMF are expected to be 4--6 times higher
 for most of our Galaxy (\cite{Becketal}), the cosmic ray
 spectrum extends to energies higher by  at least 2 and a half
 orders of magnitude. This creates an opportunity to study the
 large scale structure of GMF through the observations of the
 arrival directions of the ultra high energy cosmic rays (UHECR),
 those of energy above 10$^{19}$ eV. 

  UHECR are observed by the air showers that they initiate when the
 primary UHECR particles interacts in the atmosphere. Ground based
 air shower arrays register the arrival of a large shower by the
 coincidental arrival of a large number of charged particles in
 distant particle detectors. Since the flux of UHECR is extremely low 
 ( 0.5 per km$^2$ per year per steradian above 10$^{19}$ eV) the
 shower arrays designed for their detection are by necessity
 very sparse and the detection yields only the energy and 
 the arrival direction of the primary UHECR. The nature of
 that particle could be only derived from a large enough statistical
 sample.

  Although the origin, and even the nature, of the cosmic rays
 with such high energy is yet unknown, it is easy to understand
 their value for revealing the general structure of GMF in the
 natural assumption that they are mostly ionized hydrogen atoms
 (protons) and are of extragalactic origin. Protons of energy
 above 10$^{19}$ eV do not suffer significant energy loss on
 galactic scalelengths. They are only deflected in magnetic fields
 extending on scale larger than 1 kpc. If their arrival distribution
 on their entry into our Galaxy is not strictly isotropic, GMF would
 have a focusing (or defocusing) effect on their arrival distribution
 at Earth, which will reflect the general GMF structure, rather then
 the local magnetic field in the vicinity of the solar system.
 This is of course only possible if the UHECR are nuclei of
 astrophysical origin (\cite{Hillas,Wolfy,RachBier}), rather than
 gamma rays generated in exotic processes, such as topological
 defects (\cite{Bhat,Gill}).

  The current experimental statistics of such events is not
 sufficiently large to make definite  conclusions on their
 arrival directions, although some potentially significant
 anisotropies have been recently observed by two independent
 groups (\cite{Stanevetal,akeno}). The Auger
 project (\cite{Auger}), however, proposes  building
 of two identical air  shower arrays, situated in the Northern and
 in the Southern hemispheres in such a way that they provide all sky
 coverage. In few years of observation these detectors would increase
 the world UHECR statistics by more than a factor of 10.
  
  As we will show further down, such much improved experimental
 statistics will contribute not only to the important question
 of the origin of the UHECR, but also to our knowledge of 
 the large scale structure of the galactic magnetic field.
 This article is organized in the following way: in section 2
 we introduce the galactic magnetic field models used in this
 study. Section 3 gives the method of proton propagation in
 GMF models and the general results. Section 4 shows how
 the arrival directions of an experimental sample of UHECR is    
 affected by the chosen GMF models and gives the general
 conclusions from this study.

\section{The galactic magnetic field models.}

 We have build two galactic magnetic field models, which implement
 the general ideas of the large scale structure of the galactic 
 magnetic field and incorporate the knowledge stemming from
 experimental observations. Observations of our Galaxy, as well of 
 many other galaxies (\cite{Becketal,Kronberg}),
 show that the regular component of the galactic magnetic fields
 could be well described by spiral fields with 2$\pi$ (axisymmetric,
 ASS) or $\pi$ (bisymmetric, BSS) symmetry. In $z$--direction,
 i.e. perpendicular the the galactic plane, the fields are either
 of odd (dipole type) or even (quadrupole type) parity, A and S type
 respectively (\cite{Becketal}). There are also observations
 of magnetic fields of mixed type.
 
  As we show further down, these four general types of magnetic field
 structure would affect the propagation of UHECR in different ways.
 We have therefore made
 two extreme combinations and build 1). a bisymmetric field model
 with field reversals and odd parity (BSS\_A model); 
 and 2). an axisymmetric field model without field reversals and
 with even parity (ASS\_S model). Both models incorporate the
 following observational knowledge:\\
  1). The solar system is at a galactocentric distance $R$ = 8.5 kpc. 
 The local magnetic field in the vicinity of the solar system
 has a strength of $\sim$2$\mu$G in direction of $l=$90$^\circ$.
 The GMF pitch angle $p$ was taken to be -10$^\circ$~(\cite{RandLyne}).\\ 
  2). There is a reversal of the magnetic field (in the BSS\_A model)
 at $l$=0$^\circ$ at distance of 0.5 kpc and a second reversal at
 distance of $\sim$3 kpc~(\cite{RandLyne,Val1991}).
 In the ASS\_S model there are no reversals  and the magnetic field
 strength decreases to zero at the same locations.\\

  The field strength at a point $(r,\theta)$ in the galactic plane is 
 \begin{equation}
 B(r,\theta)\;=\;B_0(r) \cos (\theta - \beta \ln{{r}\over{r_0}})\;,
 \end{equation} 
 for the bisymmetric model, where $r_0$ is the galactocentric distance
 of the location with maximum field strength at $l$=0$^\circ$ and
 $\beta = 1/\tan{p}$ = --5.67. In our representation $r_\circ$ = 10.55 kpc.
 The BSS\_A model is very much like the model of~(\cite{hanqiao})
 scaled to a galactocentric distance of 8.5 kpc.
 In the ASS\_S model
 \begin{equation}
 B(r,\theta)\;=\;B_0(r) |\cos (\theta - \beta \ln{{r}\over{r_0}})|\;,
 \end{equation} 
 The $\theta$ and $r$ components of the field are correspondingly
 \begin{equation}
 B_\theta\;=\;B(r,\theta) \cos{p}, \; \; B_r\;=\;B(r,\theta) \sin{p}\;.
 \end{equation} 
 There is no $z$ component in the basic models. $B_0(r)$ is taken to
 be 3$R/r\;\mu$G as in~(\cite{Sofue}), i.e.
 6.4 $\mu$G at r = 4 kpc and
 constant at that value in the central region of the Galaxy. This radial
 dependence is consistent with the field strengths inferred from
 pulsar rotation measures (\cite{RandLyne}). In our models the field
 extends to galactocentric distances of 20 kpc in all directions.

\begin{figure}
\plotone{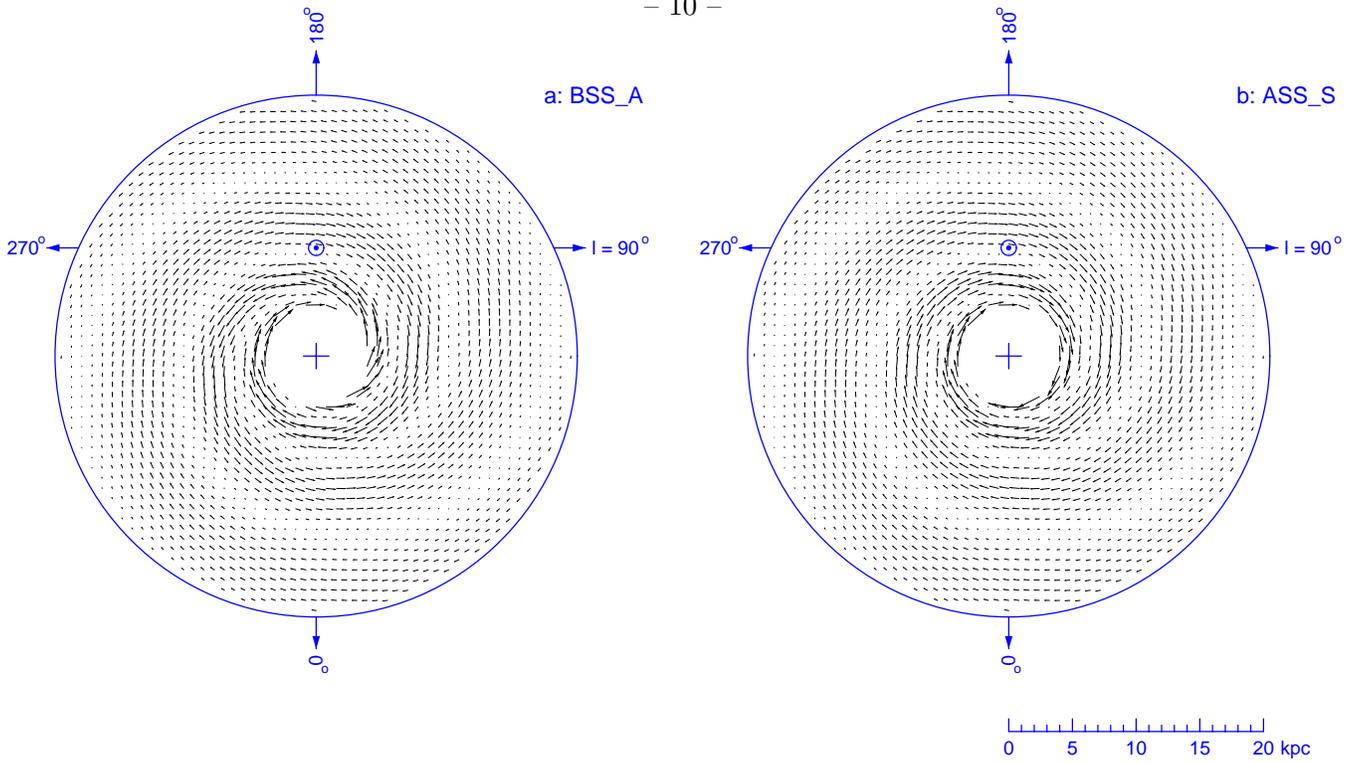}
\caption{ The direction and strength of the regular magnetic field in the
 galactic plane is represented by the length and direction of the arrows
 for the bisymmetric odd parity model BSS\_A (a) and the axisymmetric model
 with even parity ASS\_S (b). The field inside the galactocentric circle
 of radius 4 kpc follows the general structure of the models with
 $B_0(r)\;=\;B_0({\rm 4 kpc})$.}
\label{fig1}
\end{figure}

\placefigure{fig1}
 
  The size and field strength in the galactic halo is extremely important
 for the proton trajectories. We have assumed an extended
 halo with two scaleheights. The field strengths above and below
 the galactic plane are
 \begin{equation}
 |B(r,\theta,z)|\;=\;|B(r,\theta)| \exp{(z/z_0)}
 \end{equation}
 with $z_0$ = 1 kpc for $|z|<$ 0.5 kpc and $z_0$ = 4 kpc for $|z| >$ 0.5 kpc.
 The odd parity (BSS\_A) model preserves the field direction at the disk 
 crossing while the even parity (ASS\_S) model changes it. 
 
  Fig.~1 shows the field strength and direction in the galactic plane
 for both models.

 \section{ Proton propagation in the galactic magnetic field}

 To follow the particle trajectory in the GMF we use a well known
 technique which is standard in determining the geomagnetic 
 cutoffs for low energy cosmic rays and their asymptotic 
 directions -- backtracking antiprotons (\cite{Fluck,Bieberetal})
 injected at Earth in
 different directions. The negatively charged antiprotons injected
 in certain direction at Earth will follow exactly the same trajectory
 as a positively charged proton arriving at Earth from the same
 direction. So we inject antiprotons at different galactic longitude
 $b$ and latitude $l$ at the location of the solar system and follow
 their propagation in the GMF models
 until they reach a distance of 20 kpc from the galactic center.
 Then we calculate the `true' values of ($l,b$), $l_{true}$ and
 $b_{true}$, from which the particle reached the Galaxy before its
 direction was changed by the GMF. We backtracked the antiprotons
 by integrating the equations of motion of charged particles using
 an Runge--Kutta  method  with adaptive stepsize control
 (\cite{Lipari}). 

  Fig.~2 shows the results for the two GMF models. Protons of energy
 100, 80, 60, 40, and 20 EeV were injected at Earth at galactic
 latitudes from $b$ = -75$^\circ$ to $b$ = +75$^\circ$ and longitudes
 from $l$ = 45$^\circ$ to $l$ = 315$^\circ$. The deflection for a total
 of 304 injection directions were calculated. The region of the galactic
 center was intentionally omitted from the calculation because of the
 very large uncertainty in the GMF magnitude and structure. 
 Each of the shown trajectories shows the deflection suffered by a proton
 as a function of its energy, as indicated on the legend to Fig.~2. The origin
 of each trajectory is the injection direction, and the arrow ends in
 the direction at which  a 20 EeV proton would arrive at the galactic
 boundary (20 kpc from the galactic center)  to be observed as coming from
 the injection  direction at Earth. 

\begin{figure}
\plotone{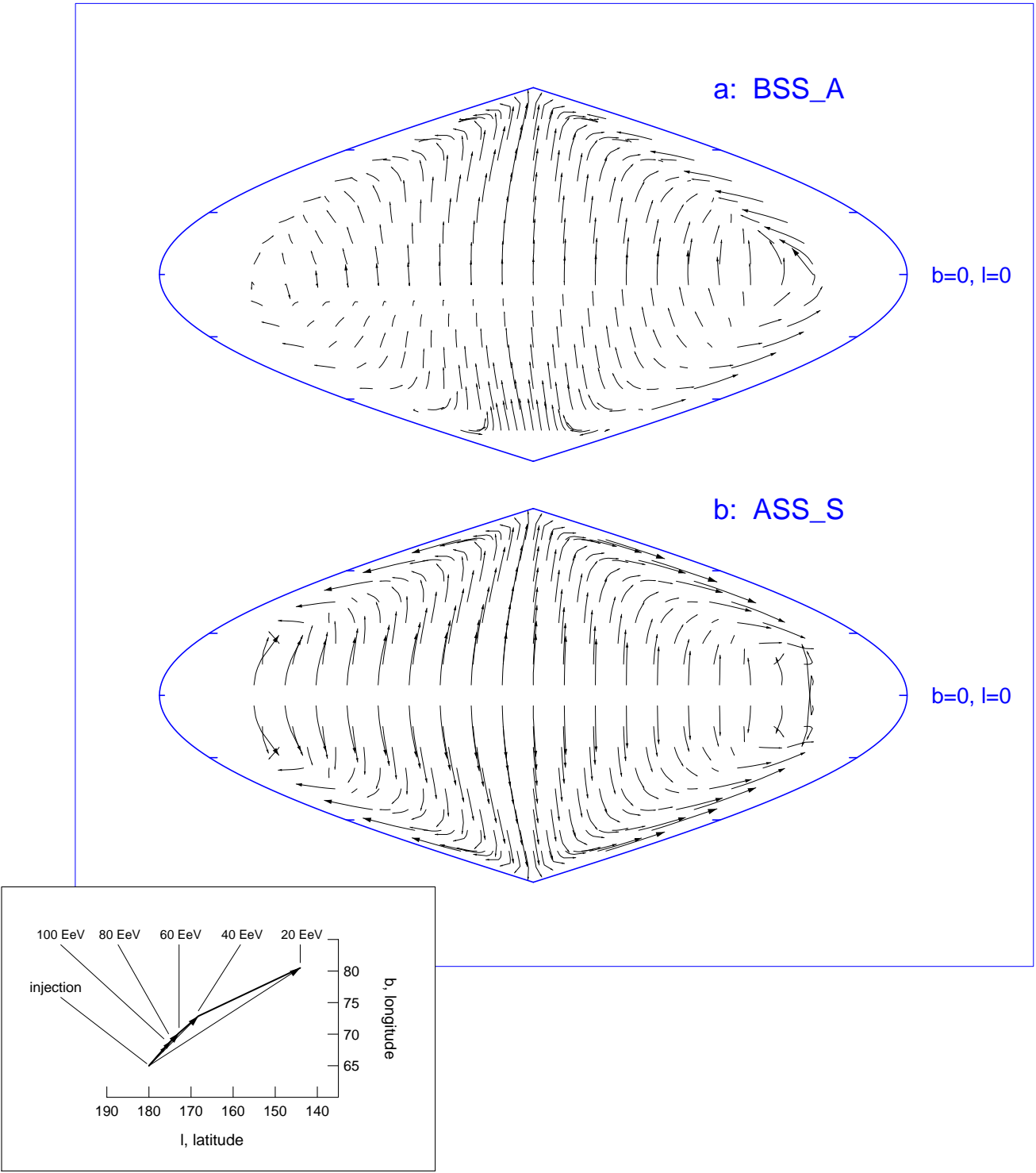}
\caption{ Deflection of protons of energy 20, 40, 60,
 80, and 100 EeV in the galactic magnetic field for the BSS\_A (a)
 and ASS\_S (b) models. The legend in the lower left corner explains
 how each trajectory is plotted. The particle is injected at Earth in
 the position of the origin of the plotted trajectory. A proton of
 energy 100 EeV would move to the angle of the first arrow in its
 propagation through the Galaxy, a proton of energy 80 EeV would move
 to the angle of the next arrow, etc. while a proton of energy 20 EeV 
 would move to the tip of the arrow in the main figure. The figure is
 centered on the galactic anticenter. }
\label{fig2}
\end{figure}

\placefigure{fig2}

  Let us first examine the proton trajectories for model BSS\_A. There is
 a general trend for a flow from North to South, which is quite strong
 in the direction of the galactic anticenter. There is also a general
 trend for a flow in the direction away from the galactic center for
 longitudes $b$ less than about -20$^\circ$, and a flow towards the galactic
 center at the corresponding Northern longitudes. One has to be
 reminded  that arrows in Fig.~2 show the reverse trajectory -- a particle
 arriving at the Galaxy from the position of the tip of the arrow
 would be observed at Earth at its origin. The RMS deflection
 angles ($\Delta\Psi_{RMS}$) are shown in Table~1 as a function of
 the proton energy. As could be seen from Fig.~2 deflection angles
 are a strong function of the injection direction. The variations
 in the magnitude of the deflection angle are large and the
 average deflection angle $<\Delta\Psi>$ is always smaller than
 the RMS value.  $\Delta\Psi_{RMS}$ values could only be taken as
 a guide for the magnitude of the expected deflection and its energy
 dependence, which is slightly stronger than linear for energies
 below 40 EeV.

\placetable{tbl1}

  Fig.~2b (model ASS\_S) shows quite a different picture. To start with,
 the deflections are quite a bit stronger. The main reason is that in the
 axisymmetric model protons always propagate through magnetic field of the
 same polarity (except in the equatorial region), while in the
 bisymmetric model they move from regions of positive polarity into
 regions of negative polarity and {\em vice versa} where they are
 respectively deflected in opposite directions. Because of this
 the net deflection in a bisymmetric field is smaller. The most
 crucial difference is caused by the even parity of the ASS\_S model --
 there is a very strong flow of protons towards the galactic plane,
 especially at longitudes between 90$^\circ$ and 270$^\circ$.
 This would make extragalactic cosmic rays appear as if they are
 actually arriving from the direction of the galactic plane. At high
 positive and negative longitudes there is now a flow away from
 $l$ = 0$^\circ$. 

  The very large deflections at extreme Northern and Southern
 longitudes may not be realistic because of the possibly
 exaggerated size and strength of the galactic halo in our
 magnetic field models. These are still appropriate for this
 exercise which is used here to demonstrate the effect of the
 galactic magnetic field on the particle propagation
 in the Galaxy. 

  A possible $z$--component of the galactic magnetic field, that
 would correspond to the existence of galactic wind, would also
 have extreme effects on the particle deflection, as shown in Fig.~3,
 which follows the propagation of protons in the two field models
 which now have a constant $B_z$ component of 0.3 $\mu$G. In the
 BSS\_A model $B_z$ is always directed to North, while in the
 ASS\_S model $B_z$ points at North in the Northern galactic 
 hemisphere and changes direction at the crossing of the galactic
 plane. 

\begin{figure}
\plotone{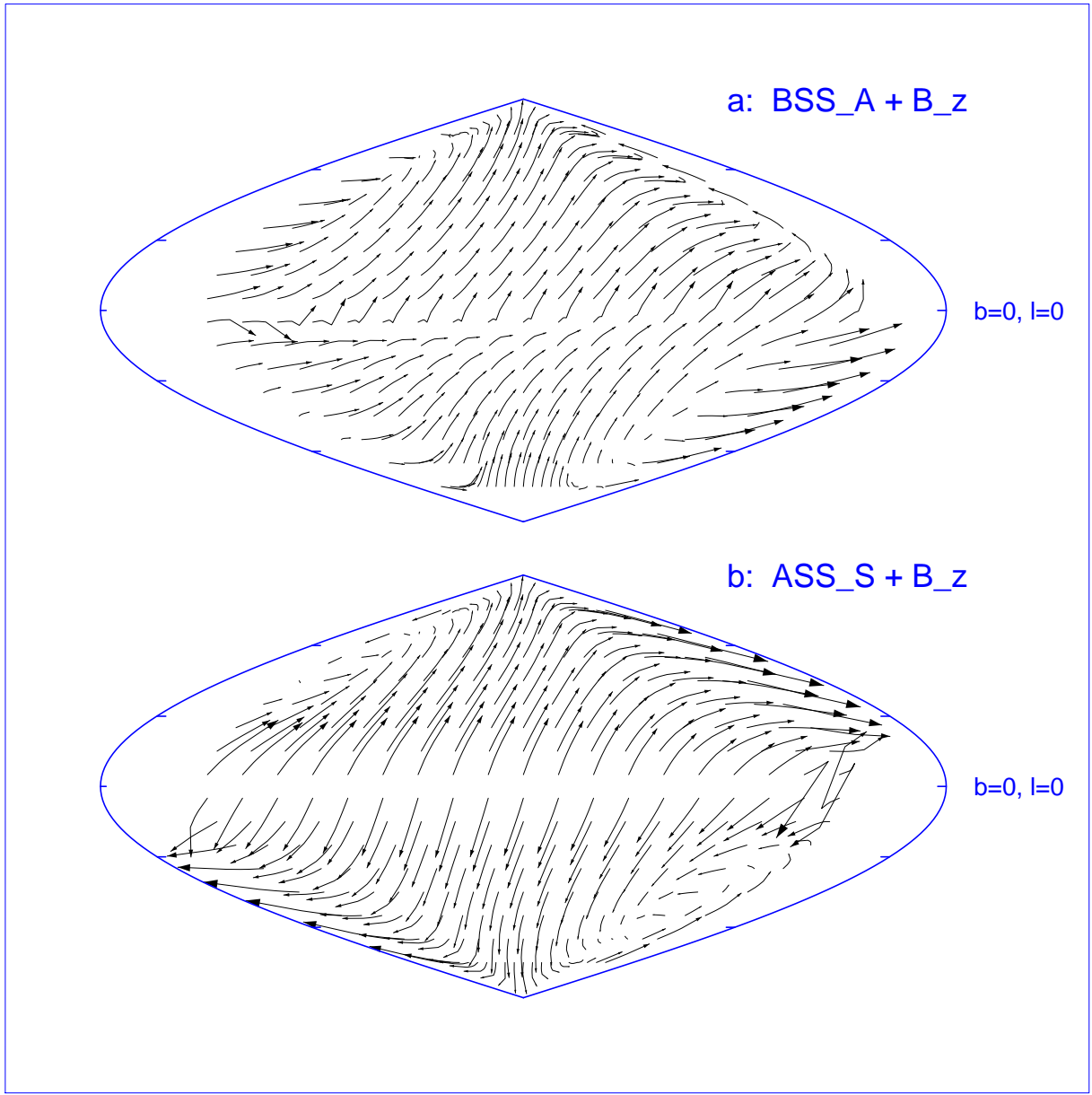}
\caption{ Same as Fig.~2 for models with a constant
 $B_z$ component with strength of 0.3 $\mu$G. In the BSS\_A case
 $B_z$ always points North, while in the ASS\_S case it changes direction
 at the crossing of the galactic plane.}
\label{fig3}
\end{figure}

\placefigure{fig3}

  The existence of a $B_z$ component makes the deflection pattern
 much more complicated, especially in the ASS\_S model. Particles
 appear to change their deflection from positive to negative
 as a function of their rigidity. These could be recognized as
 trajectories in Fig.~3 that cross the equatorial plane. The asymmetry
 in the deflection patterns becomes stronger, and protons injected
 at very high longitudes are affected the most. Fig.~3 demonstrates
 how a relatively small, hardly observable, $B_z$ component could
 change the deflection pattern for UHECR in the Galaxy. A more
 realistic $B_z$ component, related to the field strength and
 direction in the galactic plane, would complicate the deflection
 pattern even more. 

  These complications, however, concern mostly the details of the
 proton propagation in the Galaxy. The main features observable in Fig.~2
 are still present in the case of non--zero $B_z$ component. The general
 flows toward the galactic plane (BSS\_A) and towards and away from
 $l$ = 0$^\circ$ are somewhat modified, but still the major
 propagation effect. 

 \section{Discussion and conclusions}

  The magnitude of the deflections in the galactic magnetic field
 requires that the arrival direction of UHECR are corrected for
 deflection in GMF before their arrival directions are compared
 to locations of powerful astrophysical sources. As an example 
 we perform this correction in Fig.~4 for the experimental sample
 that was used in~(\cite{Stanevetal}) in search of 
 correlations  with the the galactic and the
 supergalactic(\cite{Vauc56,Vauc76}) planes.
 These events come from four air shower arrays that have
 operated in the Northern hemisphere at different times since 1960's:
 Volcano Ranch (\cite{Linsley,Lawrence,Yakutsk,AGASA}).
 The total number of events above 2$\times$10$^{19}$ eV is 143.

  It is interesting to 
 observe that the correction in either of the GMF models did not
 change appreciably the RMS distance of the UHECR direction with
 respect to the supergalactic plane $b_{RMS}^{SG}$. The RMS distance
 to the galactic plane in the ASS\_S model was increased by several
 degrees from its original (already unlikely high) value for UHECR
 of energy above 4$\times$10$^{19}$ eV. The distribution about
 the two planes, however, is now quite a bit different. A closer
 examination of Fig.~4 reveals several interesting phenomena, that are
 not strong enough to draw significant conclusions, but
 are still interesting. The effects are stronger in Fig.~4b, which
 corresponds to the ASS\_S model.

  There are several groups of experimental events that seem to
 be significantly closer to the supergalactic plane after the
 correction for the deviation in the GMF. One such group
 could be observed at northern longitudes ($b>$60$^\circ$) at
 $l$ = 180$^\circ$ to 200$^\circ$. The correction made all these
 showers appear as coming from the Virgo cluster. Another
 group of event that move closer to the supergalactic plane are
 positioned at $b$ = 0$^\circ$ to 30$^\circ$ and $l$ = 140$^\circ$ to
 160$^\circ$. Those two groups of UHECR could be observed in both
 models. Quite an interesting development appears in a wide range
 south of the galactic plane ($b$ from -5$^\circ$ to -50$^\circ$)
 at $l$ = 135$^\circ$ to 160$^\circ$. Many UHECR in this region
 land almost exactly on the supergalactic plane after the correction
 for deflection in the ASS\_S GMF model. The effect does not exist 
 in the BSS\_A model where the correction takes the UHECR directions
 to the North.  
 
  Events that belong to any of these three groups appear more correlated 
 with the supergalactic plane than before the deflection correction.
 On the other hand, there is a vast region of the sky ($b>$0$^\circ$,
 $l<$130$^\circ$, where the corrected UHECR positions are further away
 from the supergalactic plane. This is the most general trend which can
 be observed in Fig.~4, which is also essential for net zero result
 of the correction on $b_{RMS}^{SG}$. The correction also seems to
 have produced a focusing effect for several UHECR in the region
 $b$ from 5$^\circ$ to 15$^\circ$, $l$ from 140$^\circ$ to 160$^\circ$,
 which now come together at angles comparable to the angular
 resolution of the air shower arrays.
  
  Fig.~4 is introduced here only as an example for the propagation
 effects in the Galaxy. Although the corrected arrival direction
 maps should be inspected for association with powerful astrophysical
 objects, one cannot expect to draw any major conclusions on the basis
 of this statistically limited sample in view of the uncertainties
 in the galactic magnetic field models. The total world statistics would
 quickly increase by  factors of 10 to 100 when the Auger project
 comes into operation. Auger proposes two giant air shower arrays with
 area of 3000 km$^2$ each in both hemisphere. For comparison the biggest
 operational air shower array (AGASA) has an effective area of 200 km$^2$. 
 After few years of operation, Auger will provide an UHECR statistics
 of tens of thousands. At this stage one could, and should, study
 the deflections of  charged particles in the galactic magnetic field
 with the dual purpose of establishing the UHECR origin and
 the large scale structure of the magnetic field. 

  The magnetic field studies would be most fruitful if we are lucky enough
 to see individual sources of charged UHECR. Then we would be able 
 to see the highest energy particles coming directly from the source
 with minimal deflection and lower energy ones creating a halo around
 them. Such a picture, although not with high enough statistical
 significance, is suggested by the study of~(\cite{Stanevetal}).
 With much higher statistics one should be able to follow the deviation
 in the GMF as a function of the UHECR energy and derive much more
 accurate information of the magnetic field structure. 

  Even now, however, the research presented here helps to understand
 some previously observed phenomena. In (\cite{Stanevetal})
 a sudden and strong decrease of the average distance of the UHECR
 directions to the supergalactic plane was found between energies of
 20 and 40 EeV. A brief inspection of Table~1 indicates that this
 is the energy range where the deflection in the galactic magnetic
 field becomes comparable to the thickness of the supergalactic
 plane, and that one would expect a strong decrease of any
 correlation at energy of 20 EeV. The average deflection angles 
 also indicate the scale of clustering of UHECR around their
 potential sources. Another practical application of the presented
 technique is an analysis of the apparent clustering  of the UHECR
 events detected by the Souther hemisphere SUGAR air shower
 array~(\cite{Roger}).

  The strength and aspect of the UHECR deflection depends strongly
 on the viewing area of the particular experiment, which are usually
 of order 1 steradian.  It is thus very interesting to see if the
 observed shift between the arrival directions of the Haverah Park
 (\cite{Stanevetal}) and AGASA (\cite{akeno}) events
 in respect with the supergalactic plane could be explained by the
 different viewing areas of the two experiments, which are at
 latitudes of 54$^\circ$N and 35.5$^\circ$N respectively. The shift
 could also be enhanced by a difference in the energy normalization
 of the two experiments.

  The current research does not take into account the possible
 deflection of charged UHECR in the extragalactic magnetic fields.
 This would complicate the picture, since the extragalactic fields
 could be strong enough  to cause even bigger deflections on the
 Mpc extragalactic  scalelengths. Kronberg (1994) already
 suggested to use future UHECR statistics for studies of the
 extragalactic magnetic fields. The same procedure as the one
 applied here could be used, at least for some regions of the
 universe, where the information about the extragalactic magnetic
 fields is relatively rich, such as the Virgo cluster (\cite{Val1993}).
 In any case we have to account for the propagation in our own
 Galaxy before we go any further. 

\begin{acknowledgements} 
   The author is grateful to R.~Beck, P.L.~Biermann,  R.J.~Protheroe
 J.P.~Rachen, D.~Seckel and Z.P.~Zank for helpful discussions and
 to M.~Nagano and the AGASA group for sharing results prior to
 publication. This work is supported in part by the U.S. Department
 of Energy under contract  DE--FG--91ER40626. 
\end{acknowledgements}

\begin{table}
\begin{center}
\caption{ RMS deflection angle  ($\Delta\Psi_{RMS}$)
 as a function of the proton energy in EeV for the models presented
 in Figs.~2 \& 3.}
\begin{tabular}{l| c c c c c}
\hline\noalign{\smallskip}
E, EeV & 100 & 80 & 60 & 40 & 20 \\
\noalign{\smallskip}
 & \multicolumn{5}{|c}{$\Delta\Psi_{RMS}$}\\
\hline\noalign{\smallskip} 
BSS\_A  & 3.1 & 3.9 & 5.2 &  7.9 & 17.7 \\
ASS\_S  & 4.0 & 5.1 & 6.8 & 10.5 & 23.7 \\
ASS\_S\_Z  & 4.5 & 5.7 & 7.6 & 11.8 & 25.9 \\
\noalign{\smallskip}
\hline
\end{tabular}
\end{center}
\label{tbl1}
\end{table}
\newpage

\newpage
\newpage
\newpage
\begin{figure}
\plotone{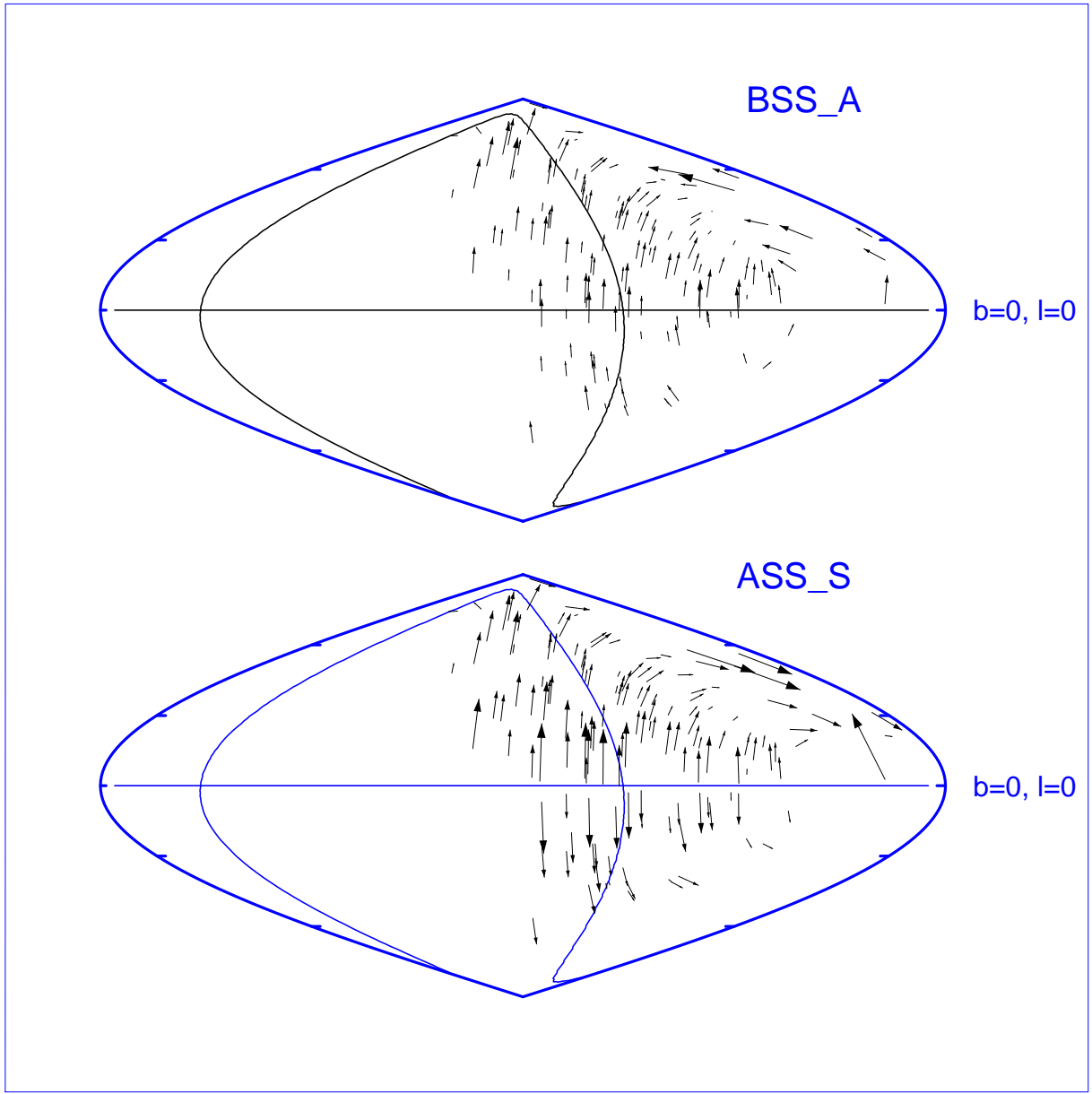}
\caption{Corrections for the arrival directions of the
 experimental sample used by Stanev et al (1995) in the two magnetic
 field models. Every experimental event was tracked back from its
 detection position (origin of trajectory shown) to determine its
 direction at its arrival in our Galaxy (tip of arrow).}
\label{fig4}
\end{figure}  
\end{document}